\documentclass[twocolumn,aps,prl,showpacs,superscriptaddress,reprint]{revtex4-1}
\usepackage{graphicx}
\usepackage{amsbsy,amssymb,amsmath,bm,ulem}
\DeclareMathOperator{\atanh}{atanh}

\begin{document}

\title{Oscillatory regimes of the thermomagnetic instability in superconducting films}

\author{J. I. Vestg{\aa}rden}
\affiliation{Department of Physics, University of Oslo, P. O. Box
1048 Blindern, 0316 Oslo, Norway}
\affiliation{Norwegian Defence Research Establishment (FFI), Kjeller, Norway}
\author{Y. M. Galperin}
\affiliation{Department of Physics, University of Oslo, P. O. Box
1048 Blindern, 0316 Oslo, Norway}
\affiliation{Ioffe Physical Technical Institute, 26 Polytekhnicheskaya,
St Petersburg 194021, Russian Federation}
\author{T. H. Johansen}
\affiliation{Department of Physics, University of Oslo, P. O. Box
1048 Blindern, 0316 Oslo, Norway}
\affiliation{Institute for Superconducting and Electronic Materials,
University of Wollongong, Innovation Campus, Squires Way,
North Wollongong NSW 2500,
Australia}

\begin{abstract}
The stability of superconducting films with respect
to oscillatory precursor modes for thermomagnetic avalanches is investigated 
theoretically. The results for the 
onset threshold  show that previous treatments of
non-oscillatory modes have predicted much higher thresholds. 
Thus, in film superconductors, oscillatory  modes are far more likely to 
cause thermomagnetic breakdown.
This explains the experimental fact that flux avalanches in film
superconductors can occur even at very small ramping rates of the applied magnetic field.
Closed expressions for the threshold magnetic field and temperature, as well oscillation frequency,
are derived for different regimes of the oscillatory thermomagnetic instability.

\end{abstract}


\maketitle

The irreversible electromagnetic properties of type-II superconductors
are commonly explained in terms of the critical current density
$j_c$, as introduced by Bean \cite{bean64}. In the corresponding 
critical state the distribution of magnetic flux is nonuniform and 
metastable.  However, since $j_c$ is a decreasing function of
temperature the metastable state can become unstable
driven by the Joule heat generated during flux motion.  
In bulk superconductors this  thermomagnetic instability gives rise to
abrupt displacement of large amounts of flux,  so-called flux 
jumps, which may cause  the entire superconductor to be heated 
to the normal state \cite{wipf67, swartz68, mints81, wipf91, rakhmanov04}. 
In some cases, pronounced oscillations in magnetization and temperature have been
detected prior to such jumps \cite{legrand93, mints96-2}.

\begin{figure}[b]
  \centering \includegraphics[width=6.0cm]{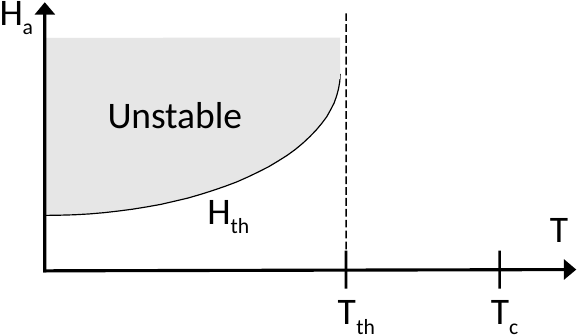}
  \caption{
    Generic thermomagnetic stability diagram of film superconductors. 
    \label{fig:HT} 
  }
\end{figure}

In film superconductors experiencing an increasing  transverse magnetic fields, 
the thermomagnetic instability gives rise to abrupt flux entry in the form of  
dendritic structures rooted at the sample edge \cite{wertheimer67}.
Using magneto-optical imaging \cite{jooss02} the residual flux distribution 
left in the film after such avalanche 
events have been observed in  many superconducting materials 
\cite{leiderer93, vlasko-vlasov00, johansen02, welling04}. 
The experiments also show that there 
is a threshold magnetic field, $H_{\rm th}$,  
for the onset of avalanche activity,
and that the unstable behavior is restricted to temperatures below a threshold 
value, $T_{\rm th}$, see Fig.~\ref{fig:HT}. These thresholds
have been explained on the basis of linear stability
analysis of the nonlinear and non-local equations governing the 
electrodynamics of such films 
\cite{mints96, aranson05, shantsev05,denisov05, denisov06, albrecht07}.
The theoretical works show that in order to trigger avalanches
an electrical field in the range $E =$ 30-100~mV/m is required.

Experimentally, one finds in films of many materials, e.g.,
MgB$_2$, Nb and NbN, that avalanches occur even when the magnetic
field is ramped very slowly, e.g., below 1 mT/s \cite{johansen01}, 
inducing correspondingly small $E$-fields.  
For a film placed in a magnetic field ramped at a rate of $\mu_0\dot
H_a$  the Bean model estimates \cite{brandt95} the $E$-field along the edge as
$E \sim \mu_0\dot H_aw$, where $w$ is the half-width of the film.
With a size of a few millimeters and a ramping rate of 1~mT/s
the edge field is $E \sim 1~\mu$V/m, i.e.,
several orders of magnitude below the theoretical threshold.
Hence, thermomagnetic avalanches should not occur at such ramping rates, quite
contrary to experiment.

This inconsistency led to the suggestion \cite{denisov06} that large
local $E$-fields are created by non-thermomagnetic micro-avalanches,
which in turn trigger the large and devastating \cite{baziljevich14}
events.  From a modelling viewpoint this idea poses severe challenges
since the proposed micro-avalanche scenarios \cite{field95, olson97,
  altshuler04, guillamon11} have not yet allowed evaluation of the
$E$-fields.

In the present work, earlier analyses of the onset conditions for
thermomagnetic avalanches are generalized by including modes with
complex instability increments. Thus, scenarios involving oscillatory
precursor behavior are here considered.  It is found that such modes,
depending on the material parameters, can have much lower thresholds
compared to those of the non-oscillatory ones. As a result, the
thermomagnetic instability can develop directly from the low $E$-field
background of the Bean critical state, without assuming existence of
micro-avalanches of unspecified nature.

\begin{figure}[t]
  \centering \includegraphics[width=7.7cm]{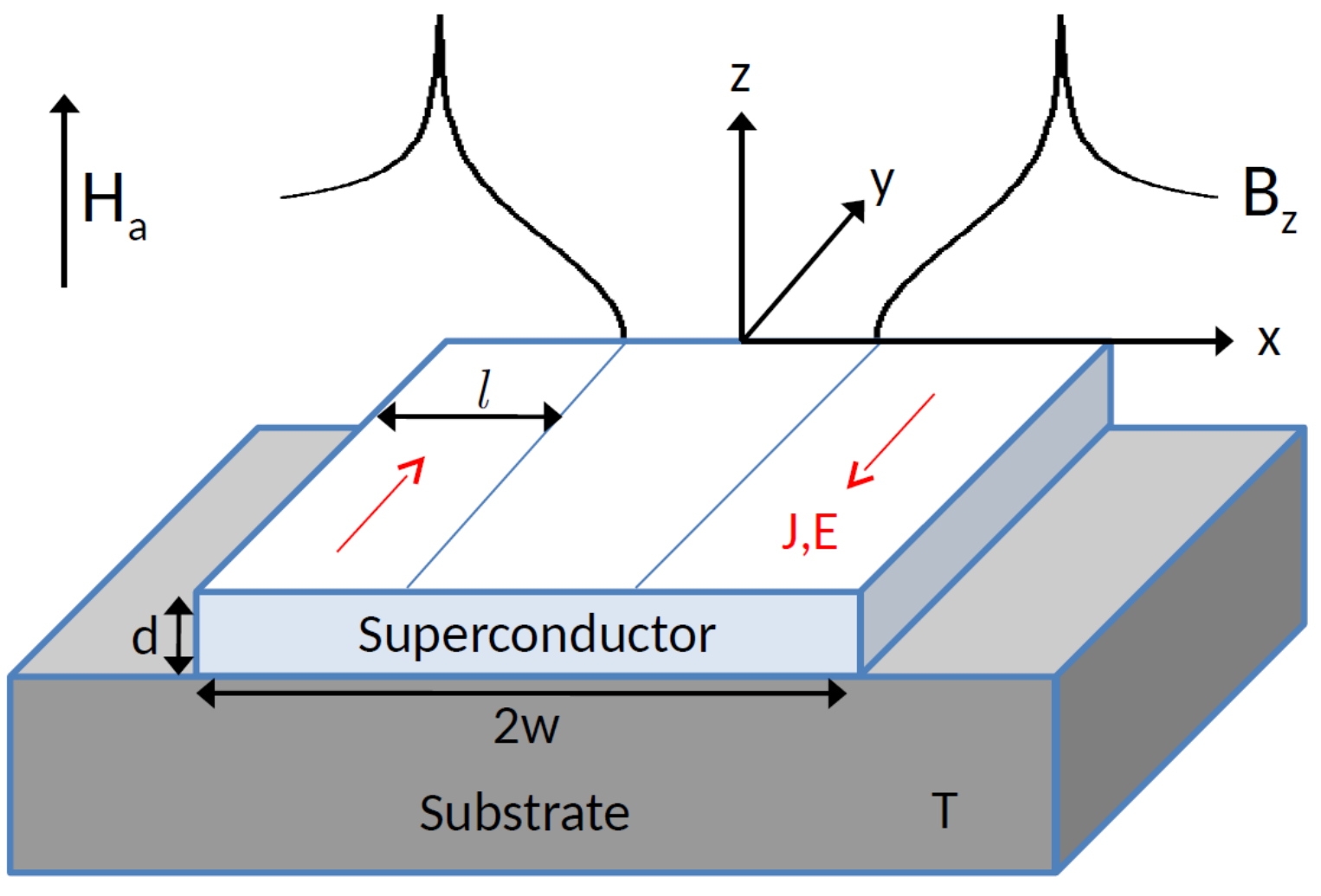}
  \caption{
    \label{fig:sample} (Color online)
     Sample geometry: A long superconducting strip in thermal contact
     with a substrate and exposed to an increasing magnetic field $H_a$
     applied along the $z$-axis, inducing currents and electrical
     fields in the $y$-direction. }
\end{figure}

Consider a superconducting film shaped as a strip of thickness $d$ and
width $2w$, where $w\gg d$. The strip is very long in the
$y$-direction, and is in thermal contact with the substrate, see
Fig.~\ref{fig:sample}. The sample is initially zero-field cooled to a
temperature, $T$, below the superconducting transition temperature,
$T_c$, whereupon a perpendicular magnetic field $H_a$ is applied at a
constant rate $\dot{H}_a$.  The overall flux dynamics in the
superconductor is assumed to follow the Bean model.  The magnetic flux
then penetrates to a depth $l$, which increases with the field as \cite{brandt93},
\begin{equation}
  \label{lx-Bean}
  l /w = 1- \cosh^{-1}(\pi H_a/dj_c)
  .
\end{equation}
This flux motion induces an electrical field,
which is maximum at the strip edge where the value is given by \cite{brandt95},
\begin{equation}
  \label{E-edge}
  E_{\rm edge} = \mu_0 \dot H_a w\tanh\left(\pi H_a/dj_c\right)
  .
\end{equation}

For perturbations of the Bean state, we describe the superconductor 
using the more general model \cite{mints96},
\begin{equation}
  \label{EJ}
  \mathbf E =
  \left\{
  \begin{array}{ll}
    \rho_n \left(J/dj_c\right)^{n-1}\mathbf J/d ,&J<dj_c \mbox{ and below}~T_c, \\
    \rho_n \mathbf J/d ,&\mbox{otherwise}.
  \end{array}
  \right.
\end{equation}
Here,  $\mathbf J$ is the sheet
current, $\rho_n$ is the normal state resistivity, and 
$n$ is the flux creep exponent. The Bean model corresponds
to the limit $n\to\infty$.

The electrodynamics follows the Maxwell equations
\begin{equation}
  \dot {\mathbf B} = -\nabla\times \mathbf E,~~ \nabla\times \mathbf H
  = \mathbf J\delta(z),~~ \nabla\cdot \mathbf B = 0,
\end{equation}
with $\nabla\cdot\mathbf J = 0$, $\mu_0\mathbf H = \mathbf B$. The
heat-flow in the strip is described by
\begin{equation}
  \label{Tdot}
  c\dot {\tilde T} =
  \kappa \nabla^2{\tilde T}- \frac{h}{d}({\tilde T}-T) +
  \frac{1}{d}\mathbf J\cdot\mathbf E
  \, ,
\end{equation}
where ${\tilde T}$ is the local temperature in the superconductor, and
$T$ is the uniform substrate temperature.  The superconductor's
specific heat is $c$, its thermal conductivity is $\kappa$, and $h$ is
the coefficient of heat transfer between the strip and the
substrate. The temperature dependencies are chosen as $c=c_0({\tilde T}/T_c)^3$, 
$\kappa=\kappa_0({\tilde T}/T_c)^3$, and $h=h_0({\tilde T}/T_c)^3$.  
For the electrodynamical parameters we use
$j_c=j_{c0}(1-{\tilde T}/T_c)$ and $n=n_0T_c/{\tilde T}$.

To determine the conditions for onset of oscillatory regimes of the
instability, consider the threshold electric field, $E_{\rm th}$,
following from linearization of Eqs.~\eqref{EJ}--\eqref{Tdot}.  As
shown in Ref. \cite{vestgarden13-metal} this gives
\begin{equation}
  \label{Eth}
  \frac{j_c}{T^*}nE_{\rm th}-\kappa k^2-\frac{h}{d}
  -\frac{2}{k}\left(k_x^2+\frac{k_y^2}{n}\right)\frac{c}{\mu_0dj_c}nE_{\rm th}
  = 0,
\end{equation}
where
$T^*=|\partial \ln j_c / \partial T|^{-1}$,
and  $k_x$, $k_y$ and $k=\sqrt{k_x^2+k_y^2}$ 
are the Fourier space wave-vectors. The instability onset is accompanied 
by temporal oscillations with frequency 
\begin{equation}
\begin{split}
    \omega^2 &= \frac{2}{k} \frac{nE_{\rm th}}{\mu_0dj_cc} \\
    \times &
    \left[
      \left(k_x^2+\frac{k_y^2}{n}\right)
      \left(\kappa k^2+\frac{h}{d}\right) 
      +\left(k_x^2-k_y^2\right)\frac{j_cE_\text{th}}{T^*}
      \right]
    . \label{omega0}
\end{split}
\end{equation}
As $H_a$ increases from zero, the most unstable modes 
correspond to $k_y=0$ and $k_x=\pi/2l$,
and in what follows only these modes are considered. 

First, at small $H_a$, the main mechanism for suppression of 
the instability is the lateral heat diffusion. 
Thus, neglecting in Eq.~(\ref{Eth}) the terms proportional to $c$ and $h$,
 one obtains 
\begin{equation}
  \label{Eth-kappa}
  E_{{\rm th}, \kappa} = \frac{\kappa T^*}{nj_c}
  \left(\frac{\pi}{2l}\right)^2 .
\end{equation}
For small $H_a$, the Eq.~\eqref{lx-Bean} gives $l\approx (w/2)(\pi H_a/dj_c)^2$, and from
Eq.~\eqref{E-edge} the electric field is $E_\text{edge}\approx \mu_0\dot H_aw\pi H_a/dj_c$. 
Inserting these expressions in Eq.~(\ref{Eth-kappa}) 
the threshold applied magnetic field becomes,
\begin{equation}
  \label{Hth}
  H_{{\rm th},\kappa} =
  \frac{dj_c}{\pi}
  \left(
  \frac{\pi^2\kappa T^*}{nw^3j_c\mu_0\dot H_a}
  \right)^{1/5} .
\end{equation}
This formula gives the threshold field as a function of temperature 
through the parameters $\kappa$, $j_c$ and $n$.
The corresponding oscillation frequency, obtained from 
Eq.~\eqref{omega0} assuming $n \gg 1$, is
\begin{equation}
  \label{omega_kappa}
 \omega_{\kappa} = \mu_0 \dot H_a n \ \sqrt{\frac{2\pi w}{\mu_0dcT^*}} \,
 ,
\end{equation}
which depends on temperature through $n$, $c$ and $T^*$. 

Then, at deeper penetration, when $l \gg (\pi/2)\sqrt{\kappa d/h}$,
the main mechanism  suppressing the 
instability is heat removal by the substrate. In this case one can  in
Eq.~(\ref{Eth}) ignore the terms
proportional to $\kappa$ and $c$, which gives, 
\begin{equation}
  \label{Eth-h}
  E_{{\rm th}, h}=\frac{hT^*}{ndj_c}
  .
\end{equation}
Combining this with Eq.~\eqref{E-edge} to eliminate the $E$-field one obtains 
the following threshold magnetic field 
\footnote{
  The low-$T$ limit
  of Eq.~\eqref{Hthh} was considered also in Ref.~\cite{mints96}.
}
\begin{equation}
  \label{Hthh}
  H_{{\rm th}, h}=
  \frac{dj_c}{\pi}\atanh\left(\frac{hT^*}{nwdj_c\mu_0\dot H_a}\right)
  .
\end{equation}
Also this case is accompanied by oscillations, and at full penetration, when
$E_{\rm edge} \approx \mu_0\dot H_aw$, the frequency is 
$\omega_{h} =  \omega_{\kappa}/\sqrt{2}$.
Thus, $\omega_{h}$ and  $ \omega_{\kappa}$ are not very different in 
magnitude, and they have a common temperature dependence.

From Eq.~\eqref{Hthh}  it follows that $H_{{\rm th}, h}$ diverges 
when the parameters satisfy the equality
\begin{equation*}
  hT^*/(nwdj_c\mu_0\dot H_a) = 1 .
\end{equation*}
When the left hand side exceeds unity the
instability will not occur, and it is therefore the condition 
determining the threshold temperature, $T_{\rm th}$.  
Thus, one finds
\begin{equation}
  \label{Tth}
  T_{\rm th}/T_c = 
  (  n_0 wdj_{c0} \mu_0\dot H_a/T_ch_0 )^{1/4} ,
\end{equation}
using the above temperature dependencies of $j_c$, $n$ and $h$.

\begin{figure}[t]
   \centering \includegraphics[width=7.5cm]{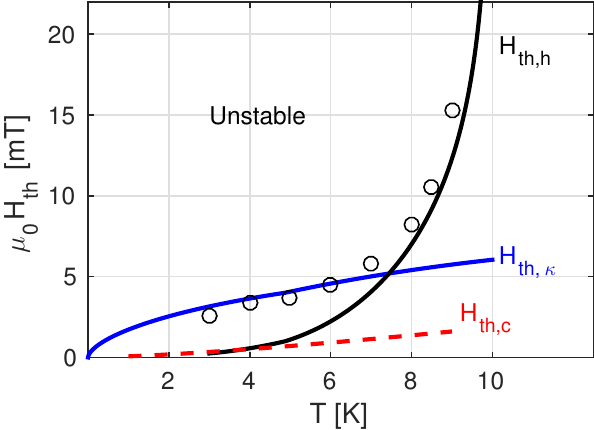}
   \caption{
     \label{fig:Hth}  (Color online)
     Threshold magnetic fields
     $H_{{\rm th}, \kappa}$ (blue), $H_{{\rm th}, h}$ (black),
     and $H_{{\rm th}, c}$ (dashed red),  as functions of temperature.
     The discrete points represent the numerical results. 
   }
\end{figure}

To verify the validity of the derived predictions near the instability onset
the set of full equations \eqref{EJ} - \eqref{Tdot} were solved 
numerically using  the procedure described in Ref.~\cite{vestgarden11}. 
Material parameters typical for 
MgB$_2$ films \cite{vestgarden11, thompson05}  were used, i.e.,  
$T_c=39~$K, $c_0=35\cdot 10^3~$J/m,
$\kappa_0=160~$W/Km$^3$, $\rho_n=7\cdot 10^{-8}~\Omega$m, $j_{c0}=
1 \cdot 10^{11}$~Am$^{-2}$, and $n_0=50$. 
The creep exponent was limited to $n=400$ at low temperatures.  
The field ramp rate was set to $\mu_0\dot H_a=600$~mT/s, 
and the sample dimensions were $w=2$~mm and $d=0.5~\mu$m.  
The substrate cooling parameter, not known from measurements, 
was taken as $h_0=1.8\cdot 10^4~$W/Km$^2$ 
to give a threshold temperature near 10 K,
in accordance with experimental observations \cite{johansen01}.
Numerical results were obtained for temperatures down to 3 K. 

\begin{figure}[t]
  \centering
  \includegraphics[width=7.7cm]{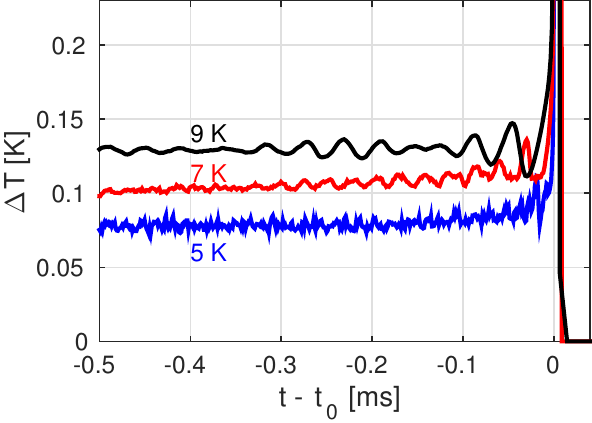}
  \caption{
    \label{fig:oscillations} (Color online)
    Temporal variations in temperature prior to avalanches 
    at 5, 7 and 9 K.
  }
\end{figure}

Figure~\ref{fig:Hth} shows  the threshold magnetic field
as function of temperature.
The full curves represent the analytical expressions 
$H_{{\rm th}, \kappa}$ and $H_{{\rm th}, h}$,  
while the discrete data show the simulation results. 
Each data point indicates the applied field when the first  dendritic avalanche 
occurred as the field increased from zero. 
At $T= 9.5$~K the simulations stopped generating  avalanches.
From the figure one sees that for $T < 7$~K, 
the graph representing $H_{{\rm th}, \kappa}$  fit the 
numerical data very well. From 7~K the data 
cross over to follow  closely the curve representing $H_{{\rm th}, h}$.
The dashed curve represents the adiabatic threshold field, $H_{{\rm th}, c}$, 
which clearly does not fit the simulations results at any temperature, see
 below for more discussion.

Direct evidence for oscillatory behavior preceding the onset of
 avalanches is presented in Fig.~\ref{fig:oscillations}. The
figure shows temporal fluctuations in the excess temperature, 
$\Delta T= \max\{\tilde T\}-T$, over an interval
of 0.5~ms prior to avalanche events at $T=$ 5, 7 and 9 K. 
The $t_0$ is the time of avalanche onset, defined as when 
$\max\{\tilde T\} = T_c$.  The graph
obtained at 9~K shows in the whole time interval clear oscillations
with one dominant frequency. During the last 0.1~ms before $ t=t_0$
the amplitude is growing significantly.  A quite similar behavior is
evident also in the graph obtained at $T= 7$~K.  The oscillations are
here smaller in amplitude, and noise is more pronounced.  Nevertheless,
nearly harmonic oscillations occur during the last 0.3~ms before
onset, and their frequency is larger than at 9~K. As in the curve for
9~K the oscillation amplitude increases towards the onset. In
addition, the dc-part of the $\Delta T $ signal also increases
slightly towards time $t_0$.  At 5~K, on the other hand, rapid fluctuations
dominate the behavior, and a characteristic frequency is not
present. The dc-part of $\Delta T $ increases also here when
approaching the time of onset.

\begin{figure}[t]
   \centering \includegraphics[width=7.6cm]{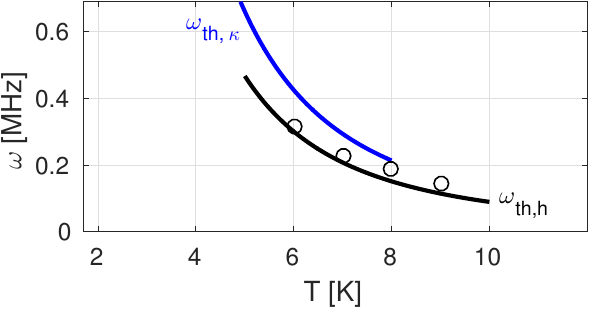}
   \caption{
     \label{fig:omega}  (Color online)
     Oscillation frequency as function of temperature.
     The discrete data represent the numerical results, 
     while the full curves are from theory.
   }
\end{figure}

The characteristic frequency, $\omega$, found from the simulation results
at temperatures between 6~K and 9~K is plotted as discrete data 
in Fig.~\ref{fig:omega}.
The $\omega$ was obtained from the location of the peak in the
Fourier spectrum of $\Delta T(t)$, 
prior to the first avalanche occurring at each temperature.
The full curves in the figure represent the analytical expressions 
$\omega_\kappa$ and $\omega_h$.
One sees that  the decrease in $\omega$ with increasing temperature is following
the curves for $\omega_h(T)$ and $\omega_{\kappa}(T)$ very well.

Regarding the adiabatic condition, i.e.,  when the 
instability is prevented only by the heat capacity of the superconductor, 
the threshold field follows from Eq.~\eqref{Eth} with $\kappa=h=0$.
Using $k_x$ with $l(H_a)$ from Eq.~\eqref{lx-Bean} in the shallow penetration limit,
one finds
\begin{equation}
  \label{Hadiabad}
  H_{{\rm th}, c} = \sqrt{\frac{2}{\pi}\frac{cT^*}{\mu_0}\frac{d}{w}} \, 
  .
\end{equation}
This expression was obtained previously \cite{shantsev05,denisov05}. New here is that also 
this threshold is associated with oscillations, 
and the frequency is contained in Eq. \eqref{omega0}.
However, for the material parameters and field ramp rate used in our simulations
the frequency curve falls far outside the scale in Fig.~\ref{fig:omega}, signalling that 
the adiabatic limit is in this case not relevant at any temperature. 
This is fully consistent also with the poor fit of the curve $ H_{{\rm th}, c}(T)$ 
to the numerical data in Fig. \ref{fig:Hth}.

Finally, it is interesting to compare the onset thresholds 
obtained for the different oscillatory regimes with those from previous works, 
where only non-oscillatory modes were considered. 
In Ref.~\cite{denisov06},
the following expression was obtained for the threshold electric field,
\begin{equation}
  E_\text{th} = \frac{T^*}{j_c}\left(\frac{\pi}{2l}\sqrt\kappa + \sqrt{\frac{h}{nd}} \right)^2
  \label{Eth-fingering}
  .
\end{equation}
By direct comparison, it follows that this threshold is higher than
the oscillatory thresholds derived in the present work.  For example,
in the case when $h=0$ the $ E_\text{th}$ in Eq.~\eqref{Eth-fingering}
is a factor $n$ larger than $E_{{\rm th}, \kappa} $ of
Eq.~\eqref{Eth-kappa}.  Thus, in an increasing applied magnetic field
the onset condition for this oscillatory regime will be met long
before that of the non-oscillatory instability.

In conclusion, several oscillatory regimes of the thermomagnetic
instability in superconducting films were analysed and explicit
onset conditions, i.e., threshold temperature, electric and  applied magnetic field, 
were obtained as functions of material parameters and field ramping rate.
The analytical work was supplemented by numerical simulations, and both
confirm that oscillatory modes are more unstable than the conventional 
ones. The results show that large-scale avalanches
can nucleate directly from the Bean critical state, rather
than being mediated by non-thermal micro-avalanches, which up to now
was the most plausible explanation for occurrence of
dendritic avalanches in films during slow field variations. 
The work provides predictions also of the characteristic oscillations frequency, 
and thus calls for new experiments investigating the nucleation mechanisms for
thermomagnetic avalanches in superconducting films.

%

\end{document}